# Polar continuum mechanics

A. R. Hadjesfandiari

G. F. Dargush

*Department of Mechanical and Aerospace Engineering*

*State University of New York at Buffalo*

*Buffalo, NY 14260 USA*

ah@buffalo.edu

gdargush@buffalo.edu

The existing polar continuum theory contains unresolved indeterminacies in the spherical part of the couple-stress tensor. This severely restricts its applicability in the study of micro and nano-scale solid and fluid mechanics and, perhaps more importantly, in the investigation of fluid turbulence phenomena, which involve a broad range of scales. In this paper, we rely on the energy equation, along with some kinematical considerations, to establish a consistent couple-stress theory for polar continua that resolves all indeterminacies. After presenting the general formulation and obtaining conservation laws, we concentrate exclusively on couple stress theory for polar fluid mechanics. We specialize the theory for linear viscous flow and consider several boundary value problems in couple-stress fluid mechanics. More generally, the resulting theory presented here may provide a basis for fundamental continuum-level studies at the finest scales.

**I. INTRODUCTION**

Classical continuum mechanics neglects the size effect of material particles within the continua. This is consistent with ignoring the rotational interaction among particles, which results in symmetry of the force-stress tensor. However, in some important cases, this cannot be true and a size dependent couple-stress theory is needed. For example, the study of micro and nano-scale solid and fluid mechanics may require a couple-stress theory. It also seems that a consistent couple-stress theory might provide a key to understanding turbulence phenomena in fluid flow.

The existence of couple-stress in materials was originally postulated in the late 19th century by Voigt[1]. However, Cosserat and Cosserat[2] were the first to develop a mathematical model to analyze materials with couple-stresses. The idea was revived and generalized much later by Toupin[3], Mindlin and Tiersten[4], Mindlin[5], Koiter[6], Nowacki[7] and other investigators in theory of elasticity. Subsequently, these formulations were brought into the domain of fluid mechanics by Stokes[8] to study the behavior of polar fluids. This development represents the simplest generalization of the classical theories of solids and fluids, which allows the presence of couple-stresses.

However, there are some difficulties with all of these formulations. Perhaps the most disturbing troubles are the indeterminacy of the spherical part of the couple-stress tensor and the appearance of the body couple in the constitutive relation for the force-stress tensor[4]. Another issue is the appearance of the body couple in the constitutive relation for the force-stress tensor[5]. Therefore, the formulation of a determinate couple-stress theory is a fundamental step in advancing continuum mechanics.

Here we develop a consistent couple-stress theory for polar continua. In Section II, we present the definitions of force-stress and couple-stress and formulate the equations of motion. Then, based on purely kinematical considerations, we suggest in Section III the mean curvature rate tensor as the corresponding compatible measure of deformation. By using the energy rate equation, we demonstrate in Section IV that in a continuum theory of couple-stress materials, body couples cannot be distinguished from an equivalent body force system. More importantly, based on resolving properly the boundary conditions, we show that the couple-stress tensor is skew-symmetric and, thus, completely determinate. This also confirms the mean curvature rate tensor as the fundamental deformation rate measure, energetically conjugate to the couple-stress tensor. Therefore, all conservation laws in consistent forms are derived. Afterwards, in Section V, we proceed to the realm of fluid mechanics, where the constitutive relations for a viscous fluid are developed and then used to derive the equations of motion for a linear fluid. The resulting theory is applied to several important steady-state flows in Section VI. Finally, we provide some concluding remarks in Section VII.

## II. STRESSES AND THE EQUATIONS OF MOTION

For a polar continuum, it is assumed that the transfer of interaction between two particles of the body through a surface element $dS$ with unit normal vector $n_i$ occurs, not only by means of a force vector $t_i^{(n)} dS$, but also as a moment vector $m_i^{(n)} dS$. Here $t_i^{(n)}$ and $m_i^{(n)}$ are force- and couple-traction vectors, respectively. Surface forces and couples are then represented by generally non-symmetric force-stress $T_{ij}$ and couple-stress $M_{ij}$ tensors, where

$$t_i^{(n)} = T_{ji} n_j \tag{1a}$$

$$m_i^{(n)} = M_{ji} n_j \tag{1b}$$

Consider an arbitrary part of a material continuum occupying a fixed control volume $V$ bounded by boundary surface $S$ as the current configuration at time $t$. The continuity equation is

$$\int_S \rho v_i n_i \, dS + \frac{\partial}{\partial t} \int_V \rho \, dV = 0 \tag{2}$$

where $\rho$ is the mass density and $v_i$ is the velocity field of the continuum in Cartesian coordinates. By applying the divergence theorem and noticing that the volume $V$ is arbitrary, we obtain the continuity equation

$$(\rho v_i)_{,i} + \frac{\partial \rho}{\partial t} = 0 \tag{3}$$

where the comma denotes differentiation with respect to the spatial (Eulerian) coordinates. This equation can also be written as

$$\frac{D\rho}{Dt} + \rho v_{i,i} = 0 \tag{4}$$

where $\frac{D}{Dt}$ is the material or substantial derivatve. It is equivalent to

$$\frac{D}{Dt} = \frac{\partial}{\partial t} + v_j \frac{\partial}{\partial x_j} \tag{5}$$

when operating on a function of space and time.

Next, assume $\psi$ is a property of the material specified per unit mass, such as energy or momentum density, in the current configuration. Therefore, the volume integral $\int_V \rho \psi dV$ over a fixed control volume $V$ gives the total energy or momentum in this control volume. The Reynolds transport theorem gives the material or substantial time derivative of this volume integral as

$$\frac{D}{Dt}\int_V \rho \psi dV = \frac{\partial}{\partial t}\int_V \rho \psi dV + \int_S \rho \psi v_i n_i dS \tag{6}$$

By applying the divergence theorem and using the continuity equation, we obtain the Reynolds transport theorem as

$$\frac{D}{Dt}\int_V \rho \psi dV = \int_V \rho \frac{D\psi}{Dt} dV \tag{7}$$

The linear and angular balance equations for this part of the body can be written

$$\int_S t_i^{(n)} dS + \int_V \rho b_i dV = \frac{D}{Dt}\int_V \rho v_i dV \tag{8a}$$

$$\int_S \left[\varepsilon_{ijk} x_j t_k^{(n)} + m_i^{(n)}\right] dS + \int_V \left[\varepsilon_{ijk} x_j \rho b_k + \rho l_i\right] dV = \frac{D}{Dt}\int_V \rho \varepsilon_{ijk} x_j v_k dV \tag{8b}$$

where $b_i$ and $l_i$ are the body force and body couple per unit mass of the continuum, respectively. Here $\varepsilon_{ijk}$ is the permutation tensor or Levi-Civita symbol. By applying the Reynolds transport theorem to these equations, we have

$$\int_S t_i^{(n)} dS + \int_V \rho b_i dV = \int_V \rho \frac{Dv_i}{Dt} dV \tag{9a}$$

$$\int_S \left[\varepsilon_{ijk} x_j t_k^{(n)} + m_i^{(n)}\right] dS + \int_V \left[\varepsilon_{ijk} x_j \rho b_k + \rho l_i\right] dV = \int_V \rho \varepsilon_{ijk} x_j \frac{Dv_k}{Dt} dV \tag{9b}$$

Now by using the divergence theorem, and noticing the arbitrariness of volume $V$, we obtain the differential form of the equations of motion, for the usual couple-stress theory, as

$$T_{ji,j} + \rho b_i = \rho \frac{Dv_i}{Dt} \tag{10a}$$

$$M_{ji,j} + \varepsilon_{ijk} T_{jk} + \rho l_i = 0 \tag{10b}$$

It will be shown shortly that the body couple density $l_i$ is not supported in polar continuum mechanics.

## III. KINEMATICS

Consider the neighboring points $P$ and $Q$ fixed by Eulerian position vectors $x_i$ and $x_i + dx_i$ in the current configuration at time $t$. The relative velocity of point $Q$ with respect to $P$ is

$$dv_i = v_{i,j} dx_j \qquad (11)$$

where $v_{i,j}$ is the velocity gradient tensor at point $P$. As we know, although this tensor is important in the analysis of the deformation rate, it is not itself a suitable measure of rate of deformation. In any case, this tensor can be decomposed into symmetric and skew-symmetric parts, such as

$$v_{i,j} = D_{ij} + \Omega_{ij} \qquad (12)$$

where

$$D_{ij} = v_{(i,j)} = \frac{1}{2}\left(v_{i,j} + v_{j,i}\right) \qquad (13a)$$

$$\Omega_{ij} = v_{[i,j]} = \frac{1}{2}\left(v_{i,j} - v_{j,i}\right) \qquad (13b)$$

Notice that here we have introduced parentheses surrounding a pair of indices to denote the symmetric part of a second order tensor, whereas square brackets are associated with the skew-symmetric part. The tensors $D_{ij}$ and $\Omega_{ij}$ represent the strain rate and angular velocity, respectively. The angular velocity or spin vector $\omega_i$ dual to the spin tensor $\Omega_{ij}$ is defined by

$$\omega_i = \frac{1}{2}\varepsilon_{ijk} v_{k,j} = \frac{1}{2}\varepsilon_{ijk}\Omega_{kj} \qquad (14a)$$

which in vectorial form is written

$$\boldsymbol{\omega} = \frac{1}{2}\nabla \times \mathbf{v} \qquad (14b)$$

This shows that

$$\nabla \bullet \boldsymbol{\omega} = \omega_{k,k} = 0 \tag{15}$$

which means that the angular velocity is source-less. This vector is related to the angular velocity tensor through

$$\varepsilon_{ijk}\omega_k = \Omega_{ji} \tag{16}$$

which shows

$$\omega_1 = -\Omega_{23}, \; \omega_2 = \Omega_{13}, \; \omega_3 = -\Omega_{12} \tag{17}$$

Based upon the above development, the relative velocity can be decomposed into

$$dv_i = dv_i^{(1)} + dv_i^{(2)} \tag{18}$$

where

$$dv_i^{(1)} = D_{ij}dx_j \tag{19a}$$

$$dv_i^{(2)} = \Omega_{ij}dx_j \tag{19b}$$

Then, $\Omega_{ij}$ is seen to generate a rigid-like rotation of element $dx_i$ about point P, where

$$du_i^{(2)}dx_i = \Omega_{ij}dx_i dx_j = 0 \tag{20}$$

Since $\Omega_{ij}$ cannot stretch the element $dx_i$, it does not appear in a tensor measuring stretch rate of deformation. Therefore, as we know, the symmetric strain rate tensor $D_{ij}$ is the suitable measure of the rate of deformation in classical theories.

In polar continuum mechanics, we expect to have an additional tensor measuring the curvature rate of the arbitrary fiber element $dx_i$. To find this tensor, we consider the field of the angular velocity vector $\omega_i$. The relative rotation of two neighboring points P and Q is given by

$$d\omega_i = \omega_{i,j}dx_j \tag{21}$$

where the tensor $\omega_{i,j}$ is the gradient of the angular velocity vector at point P. It is seen that the components $\omega_{1,1}$, $\omega_{2,2}$ and $\omega_{3,3}$ represent the torsion rate of the fibers along corresponding coordinate directions $x_1$, $x_2$ and $x_3$, respectively, at point $P$. The off-diagonal components represent the curvature rate of these fibers in planes parallel to coordinate planes. For example, $\omega_{1,2}$ is the curvature rate of a fiber element in the $x_2$-direction in a plane parallel to the $x_2 x_3$-plane, while $\omega_{2,1}$ is the curvature rate of a fiber element in the $x_1$-direction in a plane parallel to the $x_1 x_3$-plane.

The suitable measure of curvature rate must be a tensor measuring pure curvature rate of an arbitrary element $dx_i$. Therefore, in this tensor, the components $\omega_{1,1}$, $\omega_{2,2}$ and $\omega_{3,3}$ cannot appear. Consequently, we expect that the required tensor is the skew-symmetric part of $\omega_{i,j}$. By decomposing the tensor $\omega_{i,j}$ into symmetric and skew-symmetric parts, we obtain

$$\omega_{i,j} = \mathsf{X}_{ij} + \mathsf{K}_{ij} \tag{22}$$

where

$$\mathsf{X}_{ij} = \omega_{(i,j)} = \frac{1}{2}\left(\omega_{i,j} + \omega_{j,i}\right) \tag{23a}$$

$$\mathsf{K}_{ij} = \omega_{[i,j]} = \frac{1}{2}\left(\omega_{i,j} - \omega_{j,i}\right) \tag{23b}$$

The symmetric tensor $\mathsf{X}_{ij}$ results from applying the strain rate operator to the angular velocity vector, while the skew-symmetric tensor $\mathsf{K}_{ij}$ is the rotation rate of the angular velocity vector at point $P$. From (23a),

$$\mathsf{X}_{11} = \omega_{1,1}, \quad \mathsf{X}_{22} = \omega_{2,2}, \quad \mathsf{X}_{33} = \mathsf{X}_{3,3} \tag{24a-c}$$

and

$$\mathsf{X}_{12} = \mathsf{X}_{21} = \frac{1}{2}\left(\omega_{1,2} + \omega_{2,1}\right) \tag{24d}$$

$$\mathsf{X}_{23} = \mathsf{X}_{32} = \frac{1}{2}\left(\omega_{2,3} + \omega_{3,2}\right) \tag{24e}$$

$$\mathsf{X}_{13} = \mathsf{X}_{31} = \frac{1}{2}\left(\omega_{1,3} + \omega_{3,1}\right) \tag{24f}$$

From careful examination, it can be seen that the diagonal elements $X_{11}$, $X_{22}$, and $X_{33}$ represent the above mentioned pure torsion rates of fibers along the $x_1$, $x_2$ and $x_3$ directions, respectively, while $X_{12}$, $X_{23}$ and $X_{13}$ measure the rate of deviation from sphericity of deforming planes parallel to $x_1 x_2$, $x_2 x_3$ and $x_1 x_3$, respectively[9]. Furthermore, we may recognize that this symmetric $X_{ij}$ tensor must have real principal values, representing the pure twist rates along the principal directions. Thus, we refer to $X_{ij}$ as the torsion rate tensor and we expect that this tensor will not contribute as a fundamental measure of deformation rate. Instead, we anticipate that the fundamental curvature tensor is the skew-symmetric rotation of rotation tensor $K_{ij}$. This will be confirmed in the next section through consideration of couple-stresses in the energy equation.

We also may arrive at this outcome by noticing that only the part of $d\omega_i$ that is normal to element $dx_i$ produces a pure curvature rate. Therefore, by decomposing $d\omega_i$ into

$$d\omega_i = d\omega_i^{(1)} + d\omega_i^{(2)} \tag{25}$$

where

$$d\omega_i^{(1)} = X_{ij} dx_j \tag{26a}$$

$$d\omega_i^{(2)} = K_{ij} dx_j \tag{26b}$$

we notice

$$d\omega_i^{(2)} dx_i = K_{ij} dx_i dx_j = 0 \tag{27}$$

This shows that $d\omega_i^{(2)}$ is the component of $d\omega_i$ normal to $dx_i$. Therefore, the tensor $K_{ij}$ seems to be the suitable curvature rate tensor, which is represented by

$$[K_{ij}] = \begin{bmatrix} 0 & K_{12} & K_{13} \\ -K_{12} & 0 & K_{23} \\ -K_{13} & -K_{23} & 0 \end{bmatrix} \tag{28}$$

where the non-zero components of this tensor are

$$K_{12} = -K_{21} = \frac{1}{2}(\omega_{1,2} - \omega_{2,1}) \tag{29a}$$

$$K_{23} = -K_{32} = \frac{1}{2}(\omega_{2,3} - \omega_{3,2}) \tag{29b}$$

$$K_{13} = -K_{31} = \frac{1}{2}(\omega_{1,3} - \omega_{3,1}) \tag{29c}$$

Now we may recognize that $K_{12}$, $K_{23}$, and $K_{13}$ are the mean curvature rates of planes parallel to the $x_1 x_2$, $x_2 x_3$, $x_3 x_1$-planes, respectively, at point $P$. Therefore, the skew-symmetric tensor $K_{ij}$ will be referred to as the mean curvature rate tensor or simply the curvature rate tensor. The curvature rate vector $K_i$ dual to this tensor is defined by

$$K_i = \frac{1}{2}\varepsilon_{ijk}\omega_{k,j} = \frac{1}{2}\varepsilon_{ijk}K_{kj} \tag{30}$$

Thus, this axial vector is related to the mean curvature tensor through

$$K_{ji} = \varepsilon_{ijk}K_k \tag{31}$$

which shows

$$K_1 = -K_{23}, \ K_2 = K_{13}, \ K_3 = -K_{12} \tag{32a-c}$$

It is seen that the mean curvature rate vector can be expressed as

$$\mathbf{K} = \frac{1}{2}\nabla \times \boldsymbol{\omega} \tag{33}$$

This shows that $\mathbf{K}$ is the rotation of the spin vector, which can also be expressed as

$$\mathbf{K} = \frac{1}{4}\nabla \times (\nabla \times \mathbf{v}) = \frac{1}{4}\nabla(\nabla \bullet \mathbf{v}) - \frac{1}{4}\nabla^2 \mathbf{v} \tag{34a}$$

$$K_i = \frac{1}{4}v_{k,ki} - \frac{1}{4}v_{i,kk} = \frac{1}{4}v_{k,ki} - \frac{1}{4}\nabla^2 v_i \tag{34b}$$

Notice that

$$\nabla \bullet \mathbf{K} = K_{k,k} = 0 \tag{35}$$

which means the curvature rate vector is source-less, similar to the angular velocity vector $\boldsymbol{\omega}$.

As a final item relating to kinematics, we should mention that the source-less spin vector is directly related to the vorticity vector ζ via

$$\boldsymbol{\zeta} = 2\boldsymbol{\omega} = \nabla \times \mathbf{v} \tag{36}$$

Of course, in fluid mechanics applications, this vorticity vector is more popular than the angular velocity vector.

## IV. ENERGY EQUATION AND ITS CONSEQUENCES IN POLAR CONTINUA

The balance of energy based on the first law of thermodynamic is

$$\frac{D}{Dt}\int_V \frac{1}{2}\rho v_i v_i dV + \frac{D}{Dt}\int_V \rho\varepsilon dV = \int_S t_i^{(n)} v_i dS + \int_S m_i^{(n)} \omega_i dS \\ + \int_V \rho b_i v_i dV + \int_V \rho l_i \omega_i dV - \int_S q_i n_i dS + \int_V \rho h dV \tag{37}$$

where $\varepsilon$ is the internal energy per unit mass, $q_i$ is the heat flux vector and $h$ is the heat source per unit mass. This equation shows that the rate of change of total energy of system in volume $V$ is equivalent to the power of the external forces and couples, heat generated and heat input.

First, we show that the body couple cannot be distinguished from a body force in a polar continuum mechanical theory.

It is seen that the power of the body couple

$$\int_V \rho l_i \omega_i dV \tag{38}$$

in (37) is the only term in the volume that involves $\omega_i$. However, $\omega_i$ is not independent of $v_i$ in the volume $V$, because we have the relation

$$\omega_i = \frac{1}{2}\varepsilon_{ijk} v_{k,j} \tag{39}$$

Therefore, by using (39), we find

$$\rho l_i \omega_i = \frac{1}{2} \rho l_i \varepsilon_{ijk} v_{k,j} = \frac{1}{2} \left( \varepsilon_{ijk} \rho l_i v_k \right)_{,j} - \frac{1}{2} \varepsilon_{ijk} \rho l_{i,j} v_k \tag{40}$$

and, after applying the divergence theorem, the body couple power in (38) becomes

$$\int_V \rho l_i \omega_i dV = \int_V \frac{1}{2} \varepsilon_{ijk} \rho l_{k,j} v_i dV + \int_S \frac{1}{2} \varepsilon_{ijk} \rho l_j n_k v_i dS \tag{41}$$

which means that the body couple $\rho l_i$ transforms into an equivalent body force $\frac{1}{2} \varepsilon_{ijk} \rho l_{k,j}$ in the volume and a force-traction vector $\frac{1}{2} \varepsilon_{ijk} \rho l_j n_k$ on the bounding surface. This shows that in polar theory, the body couple is not distinguishable from the body force. Consequently, in couple-stress theory, we must only consider body forces. Therefore, for a proper couple-stress theory, the equations of motion become

$$T_{ji,j} + \rho b_i = \rho \frac{Dv_i}{Dt} \tag{42a}$$

$$M_{ji,j} + \varepsilon_{ijk} T_{jk} = 0 \tag{42b}$$

and the energy balance reduces to

$$\frac{D}{Dt} \int_V \frac{1}{2} \rho v_i v_i dV + \frac{D}{Dt} \int_V \rho \varepsilon dV = \int_S t_i^{(n)} v_i dS + \int_S m_i^{(n)} \omega_i dS + \int_V \rho b_i v_i dV - \int_S q_i n_i dS + \int_V \rho h dV \tag{43}$$

Next, we investigate the fundamental character of the couple-stress tensor based on boundary conditions.

The prescribed boundary conditions on the surface of the body can be either vectors $v_i$ and $\omega_i$, or $t_i^{(n)}$ and $m_i^{(n)}$, which makes a total number of six boundary values for either case. However, this is in contrast to the number of kinematic boundary conditions that can be imposed. In particular, if components of $v_i$ are specified on the bounding surface, then the normal component of the angular velocity vector $\omega_i$ corresponding to twisting

$$\omega_i^{(n)} = \omega^{(nn)} n_i = \omega_k n_k n_i \tag{44}$$

where

$$\omega^{(nn)} = \omega_k n_k \tag{45}$$

cannot be prescribed independently. However, the tangential component of $\omega_i$ corresponding to the rate of bending, that is,

$$\omega_i^{(ns)} = \omega_i - \omega^{(nn)} n_i = \omega_i - \omega_k n_k n_i \tag{46}$$

may be specified in addition, and the number of geometric or essential boundary conditions that can be specified is therefore five.

Next, we let $m^{(nn)}$ and $m_i^{(ns)}$ represent the normal and tangential components of the surface couple vector $m_i^{(n)}$, respectively, where

$$m^{(nn)} = m_k^{(n)} n_k = M_{ji} n_i n_j \tag{47}$$

causes twisting, while

$$m_i^{(ns)} = m_i^{(n)} - m^{(nn)} n_i \tag{48}$$

is responsible for bending.

From kinematics, since $\omega^{(nn)}$ is not an independent generalized degree of freedom, its apparent corresponding generalized force must be zero. Thus, for the normal component of the surface couple vector $m_i^{(n)}$, we must enforce the condition

$$m^{(nn)} = m_k^{(n)} n_k = M_{ji} n_i n_j = 0 \quad \text{on } S \tag{49}$$

Furthermore, the boundary couple surface power in (43) becomes

$$\int_S m_i^{(n)} \omega_i dS = \int_S m_i^{(ns)} \omega_i dS = \int_S m_i^{(ns)} \omega_i^{(ns)} dS \tag{50}$$

This shows that the couple-stress theory of polar material does not support independent distributions of normal surface couple $m^{(nn)}$, and the number of mechanical boundary conditions also is five. This result was first established by Koiter[6] in the indeterminate couple-stress theory for solids.

From the above discussion, we should realize that on the bounding surface of the polar continuum, a normal couple $m^{(nn)}$ cannot be applied. By continuing this line of reasoning, we may reveal the subtle character of the couple-stress-tensor in the continuum. First, we notice that the energy equation can be written for any arbitrary volume with arbitrary surface within the body. Therefore, for any point on any arbitrary surface with unit normal $n_i$, we must have

$$m^{(nn)} = M_{ji} n_i n_j = 0 \quad \text{in} \quad V \tag{51}$$

Since $n_i n_j$ is symmetric and arbitrary in (51), $M_{ji}$ must be skew-symmetric. Thus,

$$M_{ji} = -M_{ij} \quad \text{in} \quad V \tag{52}$$

This is the fundamental property of the couple-stress tensor in polar continuum mechanics, which has not been recognized previously.

In terms of components, the couple-stress tensor now can be written as

$$[M_{ij}] = \begin{bmatrix} 0 & M_{12} & M_{13} \\ -M_{12} & 0 & M_{23} \\ -M_{13} & -M_{23} & 0 \end{bmatrix} \tag{53}$$

and one can realize that the couple-stress actually can be considered as an axial vector. This couple-stress vector $M_i$ dual to the tensor $M_{ij}$ can be defined by

$$M_i = \frac{1}{2} \varepsilon_{ijk} M_{kj} \tag{54a}$$

where we also have

$$\varepsilon_{ijk} M_k = M_{ji} \tag{54b}$$

These relations simply show

$$M_1 = -M_{23}, \quad M_2 = M_{13}, \quad M_3 = -M_{12} \tag{55}$$

It is seen that the surface couple vector can be expressed as

$$m_i^{(n)} = M_{ji} n_j = \varepsilon_{ijk} n_j M_k \tag{56}$$

which can be written in vectorial form

$$\mathbf{m}^{(n)} = \mathbf{n} \times \mathbf{M} \tag{57}$$

This obviously shows that the surface couple vector $\mathbf{m}^{(n)}$ is tangent to the surface.

Interestingly, the angular equilibrium equation (42b) can be expressed as
$$\varepsilon_{ijk}\left(M_{k,j}+T_{jk}\right)=0 \tag{58}$$
which indicates that $M_{k,j}+T_{jk}$ is symmetric. Therefore, its skew-symmetric part vanishes and
$$T_{[ji]}=-M_{[i,j]} \tag{59}$$
which produces the skew-symmetric part of the force-stress tensor in terms of the couple-stress vector. This result could have been expected on the grounds that the skew-symmetric stress tensor $T_{[ji]}$ is actually an axial vector and should depend on the axial couple-stress vector $M_i$. Therefore, it is seen that the sole duty of the angular equilibrium equation (42b) is to produce the skew-symmetric part of the force-stress tensor. This relation can be elaborated if we consider the axial vector $S_i$ dual to the skew-symmetric force-stress tensor $T_{[ji]}$, where
$$S_i=\frac{1}{2}\varepsilon_{ijk}T_{[kj]} \tag{60a}$$
which also satisfies
$$\varepsilon_{ijk}S_k=T_{[ji]} \tag{60b}$$
and therefore
$$S_1=-T_{[23]},\ \ S_2=T_{[13]},\ \ S_3=-T_{[12]} \tag{61}$$
By using the relations (60a) and (59), we obtain
$$S_i=-\frac{1}{2}\varepsilon_{ijk}M_{[j,k]}=\frac{1}{2}\varepsilon_{ijk}M_{k,j} \tag{62a}$$
which can be written in vectorial form
$$\mathbf{S}=\frac{1}{2}\nabla\times\mathbf{M} \tag{62b}$$
This also shows that
$$\nabla\bullet\mathbf{S}=0 \tag{63}$$

Now after resolving the character of couple-stress in polar continuum mechanics, we return to the energy balace equation. By applying the Reynolds transport equation for the left hand side of energy balance equation, we have

$$\int_V \rho v_i \frac{Dv_i}{Dt} dV + \int_V \rho \frac{D\varepsilon}{Dt} dV = \int_S T_{ji} v_i n_j dS + \int_S M_{ji} \omega_i n_j dS \\ + \int_V \rho b_i v_i dV - \int_S q_i n_i dS + \int_V \rho h dV \tag{64}$$

Then by using the divergence theorem, we have

$$\int_V \rho v_i \frac{Dv_i}{Dt} dV + \int_V \rho \frac{D\varepsilon}{Dt} dV = \int_V (T_{ji} v_i)_{,j} dV + \int_V (M_{ji} \omega_i)_{,j} dV \\ + \int_V \rho b_i v_i dV - \int_S q_i n_i dS + \int_V \rho h dV \tag{65}$$

which can be written as

$$\int_V \rho v_i \frac{Dv_i}{Dt} dV + \int_V \rho \frac{D\varepsilon}{Dt} dV = \int_V (T_{ji,j} v_i + T_{ji} v_{i,j}) dS + \int_V (M_{ji,j} \omega_i + M_{ji} \omega_{i,j}) dV \\ + \int_V \rho b_i v_i dV - \int_V q_{i,i} dV + \int_V \rho h dV \tag{66}$$

Now by using the equations of motion (42a) and (42b), we obtain

$$\int_V \rho \frac{D\varepsilon}{Dt} dV = \int_V T_{ji} v_{i,j} dV + \int_V M_{ji} \omega_{i,j} dV - \int_V \varepsilon_{ijk} T_{jk} \omega_i dV \\ - \int_V q_{i,i} dV + \int_V \rho h dV \tag{67}$$

By using the relation (16), we have

$$\int_V \rho \frac{D\varepsilon}{Dt} dV = \int_V T_{ji} (v_{i,j} - \Omega_{ij}) dV + \int_V M_{ji} \omega_{i,j} dV - \int_V q_{i,i} dV + \int_V \rho h dV \tag{68}$$

Applying (11), this can be written

$$\int_V \rho \frac{D\varepsilon}{Dt} dV = \int_V \left[ T_{ji} D_{ij} + M_{ji} \omega_{i,j} - q_{i,i} + \rho h \right] dV \tag{69}$$

Now by noticing the arbitrariness of volume $V$, we obtain the differential form of the energy equation for couple-stress continuum theory, as

$$\rho \frac{D\varepsilon}{Dt} = T_{ji}D_{ij} + M_{ji}\omega_{i,j} - q_{i,i} + \rho h \tag{70}$$

Since $M_{ji}$ is skew-symmetric, we have

$$M_{ji}\omega_{i,j} = M_{ji}\mathsf{K}_{ij} \tag{71}$$

which shows that the skew-symmetric mean curvature rate tensor $\mathsf{K}_{ij}$ is energetically conjugate to the skew-symmetric couple-stress tensor $M_{ji}$. This confirms our speculation expressed in Section II of $\mathsf{K}_{ij}$ as a suitable curvature rate tensor. Furthermore, the differential energy balance becomes

$$\rho \frac{D\varepsilon}{Dt} = T_{ji}D_{ij} + M_{ji}\mathsf{K}_{ij} - q_{i,i} + \rho h \tag{72}$$

The terms $T_{ji}D_{ij}$ and $M_{ji}\mathsf{K}_{ij}$ are called force-stress and couple-stress powers per unit volume.

Interestingly, by using the dual vectors of these tensors, we have

$$M_{ji}\mathsf{K}_{ij} = \varepsilon_{ijp}M_p\varepsilon_{jiq}\mathsf{K}_q = -\varepsilon_{ijp}\varepsilon_{ijq}M_p\mathsf{K}_q = -2\delta_{pq}M_p\mathsf{K}_q = -2M_i\mathsf{K}_i \tag{73}$$

which shows the conjugate relation between twice the mean curvature rate vector $-2\mathsf{K}_i$ and the couple-stress vector $M_i$. Therefore

$$\rho \frac{D\varepsilon}{Dt} = T_{ji}D_{ij} - 2M_i\mathsf{K}_i - q_{i,i} + \rho h \tag{74}$$

Since $D_{ij}$ is symmetric, this relation can be written as

$$\rho \frac{D\varepsilon}{Dt} = T_{(ji)}D_{ij} + M_{ji}\mathsf{K}_{ij} - q_{i,i} + \rho h \tag{75}$$

where

$$T_{(ji)} = \frac{1}{2}(T_{ji} + T_{ij}) \tag{76}$$

is the symmetric part of the force-stress tensor.

What we have presented so far is a continuum mechanics theory of couple-stress media, independent of the material properties. The conservation laws or general balance equations in differential form are

$$\frac{D\rho}{Dt} + \rho v_{i,i} = 0 \tag{77a}$$

$$T_{ji,j} + \rho b_i = \rho \frac{Dv_i}{Dt} \tag{77b}$$

$$T_{[ji]} = -M_{[i,j]} \tag{77c}$$

$$\rho \frac{D\varepsilon}{Dt} = T_{(ji)} D_{ij} - 2M_i \mathsf{K}_i - q_{i,i} + \rho h \tag{77d}$$

However, these equations on their own are not enough to describe the mechanical behavior of any particular material. To complete the specification of the mechanical properties of a material, we need to define the constitutive equations. In the following section, we specialize the theory for viscous polar fluids, which does not require any additional kinematical measures of deformation beyond those defined in Section III

## V. CONSTITUTIVE EQUATIONS FOR VISCOUS POLAR FLUIDS

A viscous polar fluid is a continuum in which the force and couple-stresses depend on rates of strain and mean curvature, such that

$$T_{ji} = -p(\rho,T)\delta_{ij} + \tau_{ij}(D_{mn},\mathsf{K}_m,\rho,T) \tag{78a}$$

$$M_i = M_i(D_{mn},\mathsf{K}_m,\rho,T) \tag{78b}$$

where $T$ is the temperature and $p(\rho,T)$ is the thermodynamic pressure, representing the only stress in the fluid when there is no rate of deformation. On the other hand, $\tau_{ij}(D_{mn},\mathsf{K}_m,\rho,T)$ is the part of the force-stress tensor which depends on $D_{mn}$, $\mathsf{K}_m$ and the viscosity of the fluid.

Therefore, a simple modification of Newtonian viscous fluid gives

$$T_{(ji)} = -p(\rho,T)\delta_{ij} + \lambda(\rho,T)D_{kk}\delta_{ij} + 2\mu(\rho,T)D_{ij} \tag{79a}$$

$$M_i = -8\eta(\rho,T)\mathsf{K}_i \tag{79b}$$

Note that the expression for $T_{(ji)}$ in (79a) is exactly the classical constitutive relation, where $\lambda(\rho,T)$ and $\mu(\rho,T)$ are the conventional viscosity coefficients and $p(\rho,T)$ is the above mentioned thermodynamic pressure. Thus, $\eta(\rho,T)$ is the only additional viscosity coefficient, which specifies the character of couple-stresses in the fluid. Now for further development, we assume these coefficients are constant. Therefore, we have

$$T_{(ji)} = -p\delta_{ij} + \lambda D_{kk}\delta_{ij} + 2\mu D_{ij} \tag{80a}$$

$$M_i = -8\eta K_i \tag{80b}$$

By using the relation in (34b), we obtain

$$M_i = 2\eta\left(\nabla^2 v_i - v_{k,ki}\right) \tag{81a}$$

or in vectorial form

$$\mathbf{M} = 2\eta\left[\nabla^2 \mathbf{v} - \nabla(\nabla\bullet\mathbf{v})\right] \tag{81b}$$

Furthermore, from (35), it is obvious that

$$\nabla\bullet\mathbf{M} = 0 \tag{82}$$

Additionally,

$$M_{i,j} = 2\eta\left(\nabla^2 v_{i,j} - v_{k,kij}\right) \tag{83}$$

Therefore,

$$M_{[i,j]} = \eta\nabla^2\left(v_{i,j} - v_{j,i}\right) \tag{84}$$

or

$$M_{[i,j]} = 2\eta\nabla^2\Omega_{ij} \tag{85}$$

and we obtain for the skew-symmetric part of the force-stress tensor

$$T_{[ji]} = -M_{[i,j]} = 2\eta\nabla^2\Omega_{ji} \tag{86}$$

The axial vector $S_i$ defined in (60a) as the dual to $T_{[ji]}$ now can be written as

$$S_i = -4\eta\varepsilon_{ijk}K_{k,j} \tag{87}$$

Therefore, the constitutive relation for vector $\mathbf{S}$ is

$$\mathbf{S} = -4\eta\nabla\times\mathbf{K} \tag{88}$$

which, of course, is consistent with (63).

The relation (88) can also be rewritten as

$$\mathbf{S} = -2\eta \nabla \times \nabla \times \boldsymbol{\omega} = 2\eta \nabla^2 \boldsymbol{\omega} \tag{89a}$$

or

$$\mathbf{S} = -\eta \nabla \times \nabla \times \nabla \times \mathbf{v} \tag{89b}$$

This remarkable result shows that the vector $\mathbf{S}$, corresponding to skew-symmetric part of the force-stress tensor, is proportional to the curl of curl of curl of the velocity vector $\mathbf{v}$.

By using the relations (80a) and (86), the total force-stress tensor can be written as

$$T_{ji} = -p\delta_{ij} + \lambda D_{kk}\delta_{ij} + 2\mu D_{ij} + 2\eta \nabla^2 \Omega_{ji} \tag{90a}$$

and from (80b), we also notice that

$$M_{ji} = -8\eta \mathsf{K}_{ji} = 4\eta \left( \omega_{i,j} - \omega_{j,i} \right) \tag{90b}$$

which is more useful than $M_i$ in practice.

We may notice that these constitutive relations are similar to those in the indeterminate couple-stress theory, when $\eta' = -\eta$ [8]. Here we have derived the couple-stress theory in which all former troubles with indeterminacy disappear. Interestingly, the ratio

$$\frac{\eta}{\mu} = l^2 \tag{91}$$

specifies a characteristic material length $l$, which is absent in classical fluid mechanics, but is fundamental in couple-stress theory.

From the definition (90a) of the force-stress tensor, we find that its trace is

$$T_{kk} = -3p + (3\lambda + 2\mu)D_{kk} \tag{92}$$

Meanwhile, the negative of spherical part of stress tensor is called the mechanical stress $p_m$, where

$$-p_m = \frac{1}{3}T_{kk} = -p + \left(\lambda + \frac{2}{3}\mu\right)D_{kk} \tag{93}$$

By using the continuity relation in the form

$$\frac{D\rho}{Dt} + \rho D_{k,k} = 0 \qquad (94)$$

we obtain

$$p - p_m = \left(\lambda + \frac{2}{3}\mu\right)D_{kk} = -\left(\lambda + \frac{2}{3}\mu\right)\frac{1}{\rho}\frac{D\rho}{Dt} \qquad (95)$$

Thus, for an incompressible flow, the thermodynamic pressure $p$ and mechanical pressure $p_m$ are the same. However, it is often assumed this is the case for every fluid, which is known as Stokes's assumption. This means that

$$\lambda + \frac{2}{3}\mu = 0 \qquad (96)$$

When the force-stress tensor

$$\begin{aligned}T_{ji} &= -p\delta_{ij} + \lambda D_{kk}\delta_{ij} + 2\mu D_{ij} - 2\eta\nabla^2\Omega_{ij} \\ &= -p\delta_{ij} + \lambda v_{k,k}\delta_{ij} + \mu(v_{i,j} + v_{j,i}) - \eta\nabla^2(v_{i,j} - v_{j,i})\end{aligned} \qquad (97)$$

is carried into the linear equations of motion, we obtain

$$-p_{,i} + (\lambda + \mu + \eta\nabla^2)v_{k,ki} + (\mu - \eta\nabla^2)\nabla^2 v_i + \rho b_i = \rho\frac{Dv_i}{Dt} \qquad (98)$$

which can be written in vectorial form

$$-\nabla p + (\lambda + \mu + \eta\nabla^2)\nabla(\nabla \bullet \mathbf{v}) + (\mu - \eta\nabla^2)\nabla^2\mathbf{v} + \rho\mathbf{b} = \rho\frac{D\mathbf{v}}{Dt} \qquad (99)$$

or alternatively as

$$-\nabla p + (\lambda + 2\mu)\nabla(\nabla \bullet \mathbf{v}) - (\mu - \eta\nabla^2)\nabla \times \nabla \times \mathbf{v} + \rho\mathbf{b} = \rho\frac{D\mathbf{v}}{Dt} \qquad (100)$$

This equation is exactly the same equation used by Stokes[8] based on work of Mindlin and Tiersten[4]. The independency of this equation from $\eta'$ could have been used as an indication that $\eta'$ would not be independent of $\eta$. Now, we know $\eta' = -\eta$.

By using the constitutive relations (90a) and (80b), we have

$$T_{(ji)}D_{ij} - 2M_i K_i = -pD_{kk} + \lambda D_{kk}D_{kk} + 2\mu D_{ij}D_{ij} + 16\eta K_i K_i \qquad (101)$$

Therefore, the energy balance equation becomes

$$\rho \frac{D\varepsilon}{Dt} = -pD_{kk} + \lambda D_{kk}D_{kk} + 2\mu D_{ij}D_{ij} + 16\eta \mathsf{K}_i\mathsf{K}_i - q_{i,i} + \rho h \qquad (102)$$

For an incompressible fluid, the continuity equation reduces to

$$\nabla \bullet \mathbf{v} = 0 \qquad (103)$$

Then, from (34), the mean curvature rate vector becomes

$$\mathbf{K} = -\frac{1}{4}\nabla^2 \mathbf{v} \qquad (104)$$

and

$$T_{(ji)} = -p\delta_{ij} + 2\mu D_{ij} \qquad (105a)$$

$$M_i = 2\eta \nabla^2 v_i \qquad (105b)$$

$$T_{[ji]} = 2\eta \nabla^2 \Omega_{ji} \qquad (105c)$$

which can be wrtten as

$$\mathbf{S} = 2\eta \nabla^2 \boldsymbol{\omega} \qquad (106)$$

Finally, we have

$$T_{ji} = -p\delta_{ij} + 2\mu D_{ij} + 2\eta \nabla^2 \Omega_{ji} \qquad (107a)$$

$$M_i = 2\eta \nabla^2 v_i \qquad (107b)$$

$$S_i = 2\eta \nabla^2 \omega_i \qquad (107c)$$

where

$$T_{kk} = -3p \qquad (108)$$

For the incompressible case, the equations of motion also reduce to

$$\rho \frac{D\mathbf{v}}{Dt} = -\nabla p + \mu \nabla^2 \mathbf{v} - \eta \nabla^2 \nabla^2 \mathbf{v} + \rho \mathbf{b} \qquad (109)$$

while, interestingly, the energy equation becomes

$$\rho \frac{D\varepsilon}{Dt} = \mu\left(v_{i,j}v_{i,j} + v_{i,j}v_{j,i}\right) + \eta \nabla^2 v_i \nabla^2 v_i - q_{i,i} + \rho h \qquad (110)$$

which can also be written as

$$\rho\frac{D\varepsilon}{Dt} = \mu(v_{i,j}v_{i,j} + v_{i,j}v_{j,i}) + \eta\nabla^2\mathbf{v}\bullet\nabla^2\mathbf{v} - q_{i,i} + \rho h \tag{111}$$

For an incompressible fluid with constant viscosity, when the body force is neglected, the equations of motion are

$$\rho\frac{D\mathbf{v}}{Dt} = -\nabla p + \mu\nabla^2\mathbf{v} - \eta\nabla^2\nabla^2\mathbf{v} \tag{112}$$

which can be written as

$$\rho\left(\frac{\partial\mathbf{v}}{\partial t} + \nabla\mathbf{v}\bullet\mathbf{v}\right) = -\nabla p + \mu\nabla^2\mathbf{v} - \eta\nabla^2\nabla^2\mathbf{v} \tag{113}$$

The nonlinear term on the left hand side corresponding to convective acceleration can be written as

$$\nabla\mathbf{v}\bullet\mathbf{v} = \frac{1}{2}\nabla v^2 - \mathbf{v}\times(\nabla\times\mathbf{v}) = \frac{1}{2}\nabla v^2 - \mathbf{v}\times\boldsymbol{\zeta} \tag{114}$$

where $v^2 = \mathbf{v}\bullet\mathbf{v}$. Therefore

$$\rho\left(\frac{\partial\mathbf{v}}{\partial t} + \frac{1}{2}\nabla v^2 + \boldsymbol{\zeta}\times\mathbf{v}\right) = -\nabla p + \mu\nabla^2\mathbf{v} - \eta\nabla^2\nabla^2\mathbf{v} \tag{115}$$

The vorticity equation then is obtained by taking the curl of this equation, which gives

$$\frac{D\boldsymbol{\zeta}}{Dt} = \nabla\mathbf{v}\bullet\boldsymbol{\zeta} + \nu\nabla^2\boldsymbol{\zeta} - \nu l^2\nabla^2\nabla^2\boldsymbol{\zeta} \tag{116}$$

where

$$\nu = \frac{\mu}{\rho} \tag{117}$$

is the kinematical viscosity of the fluid. This is the vorticity transport equation, which is as important as the equation of motion itself.

Interestingly, for two dimensional flows, the vorticity vector $\boldsymbol{\zeta}$ is perpendicular to the plane of flow and, therefore,

$$\frac{D\boldsymbol{\zeta}}{Dt} = \nu\nabla^2\boldsymbol{\zeta} - \nu l^2\nabla^2\nabla^2\boldsymbol{\zeta} \tag{118}$$

## VI. SOME IMPORTANT FLOWS

We now examine a few important flows with couple-stresses. Elementary problems of Poiseuille flows between parallel plates in circular pipes are first considered. Afterwards, we develop the solution for creeping flow past a sphere. In all three cases, the flows are assumed to be under steady-state conditions. Previously, solutions for these problems were developed by Stokes[8,10] using the indeterminate couple-stress formulation, based on the work of Mindlin and Tiersten[4].

### A. Poiseuille flow between two parallel plates

Consider the one-dimensional steady flow of an incompressible fluid between two parallel plates due to a pressure gradient in the flow direction. Stokes[8] has studied this flow within the framework of indeterminate couple-stress theory. Next, we demonstrate that his results are still valid in the complete theory of polar fluid mechanics with a particular set of material parameters. For convenience, we mostly follow his notation and approach. Let the centerline be oriented along the $x$-axis with the plates situated at a distance $2h$ apart in the $y$-direction. For this flow, we may consider the velocity field to be

$$v_1 = u(y), \qquad v_2 = 0, \qquad v_3 = 0 \tag{119a-c}$$

Therefore, the equation of motion is reduced to

$$\frac{d^4 u}{dy^4} - \frac{1}{l^2}\frac{d^2 u}{dy^2} = -\frac{1}{\eta}\frac{\partial p}{\partial x}, \qquad p = p(x) \tag{120a,b}$$

which requires $\dfrac{\partial p}{\partial x}$ to be a constant. The general solution to this equation is[8]

$$u(y) = A_0 + A_1 y + B_1 \cosh\left(\frac{y}{l}\right) + B_2 \sinh\left(\frac{y}{l}\right) + \frac{1}{2\mu}\left(\frac{\partial p}{\partial x}\right) y^2 \tag{121}$$

From this, we find that the non-zero components of $\omega_i$ and $\mathsf{K}_i$ are

$$\omega_z = \omega_{yx} = -\frac{1}{2}\frac{\partial u}{\partial y} = -\frac{1}{2}\left\{ A_1 + \frac{1}{l}B_1 \cosh\left(\frac{y}{l}\right) + \frac{1}{l}B_2 \sinh\left(\frac{y}{l}\right) + \frac{1}{\mu}\left(\frac{\partial p}{\partial x}\right) y \right\} \tag{122}$$

and

$$\mathsf{K}_x = \mathsf{K}_{zy} = -\mathsf{K}_{yz} = \frac{1}{2}\frac{\partial \omega_z}{\partial y} = -\frac{1}{4}\left\{\frac{1}{l^2}B_1 \cosh\left(\frac{y}{l}\right) + \frac{1}{l^2}B_2 \sinh\left(\frac{y}{l}\right) + \frac{1}{\mu}\left(\frac{\partial p}{\partial x}\right)\right\} \tag{123}$$

Therefore,

$$M_{yz} = 8\eta \mathsf{K}_{zy} = -2\eta\left\{\frac{1}{l^2}B_1\cosh\left(\frac{y}{l}\right) + \frac{1}{l^2}B_2\sinh\left(\frac{y}{l}\right) + \frac{1}{\mu}\left(\frac{\partial p}{\partial x}\right)\right\} \quad (124)$$

The boundary conditions to be satisfied are

$$u = 0, \quad M_{yz} = 0, \quad \text{at } y = \pm h \quad (125\text{a,b})$$

These requirements reduce the solution to

$$u(y) = -\frac{h^2}{2\mu}\left(\frac{\partial p}{\partial x}\right)\left[1 - \frac{y^2}{h^2} - \frac{2l^2}{h^2}\left(1 - \frac{\cosh\left(\frac{y}{l}\right)}{\cosh\left(\frac{h}{l}\right)}\right)\right] \quad (126\text{a})$$

$$\omega_z(y) = -\frac{h}{2\mu}\left(\frac{\partial p}{\partial x}\right)\left[\frac{y}{h} - \frac{l}{h}\frac{\sinh\left(\frac{y}{l}\right)}{\cosh\left(\frac{h}{l}\right)}\right] \quad (126\text{b})$$

$$M_{yz}(y) = -M_x = -2l^2\left(\frac{\partial p}{\partial x}\right)\left[1 - \frac{\cosh\left(\frac{y}{l}\right)}{\cosh\left(\frac{h}{l}\right)}\right] \quad (126\text{c})$$

$$M_{zy}(y) = +M_x = +2l^2\left(\frac{\partial p}{\partial x}\right)\left[1 - \frac{\cosh\left(\frac{y}{l}\right)}{\cosh\left(\frac{h}{l}\right)}\right] \quad (126\text{d})$$

We should note that, based on the formulation in Stokes[8], the couple-stress component $M_{zy}$ would be written as $M_{zy} = \frac{\eta'}{\eta}M_{yz}$. With $\eta' = -\eta$, as required by the present consistent theory, we recover (126d). Also, notice that the maximum couple-stress is at the centerline with

$$M_{yz}\big|_{\max} = -2l^2\left(\frac{\partial p}{\partial x}\right)\left[1 - \frac{1}{\cosh\left(\frac{h}{l}\right)}\right] \tag{127}$$

The shear stress has two parts. The symmetric part is

$$T_{(xy)} = 2\mu D_{xy} = -\left(\frac{\partial p}{\partial x}\right)\left[-y + l\frac{\sinh\left(\frac{y}{l}\right)}{\cosh\left(\frac{h}{l}\right)}\right] \tag{128}$$

Meanwhile,

$$\mathbf{S} = \frac{1}{2}\nabla \times \mathbf{M} = -\frac{1}{2}\frac{\partial M_x}{\partial y}\hat{\mathbf{z}} = l\left(\frac{\partial p}{\partial x}\right)\frac{\sinh\left(\frac{y}{l}\right)}{\cosh\left(\frac{h}{l}\right)}\hat{\mathbf{z}} = S_z\hat{\mathbf{z}} \tag{129}$$

and then the skew-symmetric part of the shear stress is

$$T_{[yx]} = S_z = l\left(\frac{\partial p}{\partial x}\right)\frac{\sinh\left(\frac{y}{l}\right)}{\cosh\left(\frac{h}{l}\right)} \tag{130}$$

Therefore, the total shear stress $T_{yx}$ is simply

$$T_{yx} = \frac{\partial p}{\partial x}y \tag{131}$$

Thus, the result for the shear stress in the flow direction $T_{yx}$ is exactly the same as in the non-polar case. However, for $T_{xy}$, we have

$$T_{xy} = \frac{\partial p}{\partial x}y - 2l\frac{\partial p}{\partial x}\frac{\sinh\left(\frac{y}{l}\right)}{\cosh\left(\frac{h}{l}\right)} \tag{132}$$

which is different from the non-polar case.

Now we calculate the flow rate. Let $Q$ be the flow rate between the plates per unit depth. Then,

$$Q = 2\int_0^h u(y)dy = -\frac{2}{3}\frac{h^3}{\mu}\left(\frac{\partial p}{\partial x}\right)\left[1 - \frac{3l^2}{h^2}\left(1 - \frac{l}{h}\tanh\frac{h}{l}\right)\right] \quad (133)$$

If there were no couple-stresses that develop, the flow rate would be that for the classical non-polar case. Thus, we have

$$Q_0 = -\frac{2}{3}\frac{h^3}{\mu}\left(\frac{\partial p}{\partial x}\right) \quad (134)$$

and, finally

$$\frac{Q}{Q_0} = 1 - \frac{3l^2}{h^2}\left(1 - \frac{l}{h}\tanh\frac{h}{l}\right) \quad (135)$$

which indicates that for non-zero $l$ the flow rate is always less than the classical result.

## B. Poiseuille flow through circular pipes

In this section, we consider the flow through a pipe with radius $R$ due to a pressure gradient $\frac{\partial p}{\partial z}$. Stokes[8] has also studied this flow in the context of indeterminate couple-stress theory. However, we demonstrate that the results are different in the compatible polar fluid mechanics by following his steps. For convenience, we choose cylindrical coordinates $(r, \theta, z)$, where the z-axis is the pipe center line. For this flow, we may consider the velocity field to be

$$v_r = 0, \quad v_\theta = 0, \quad v_z = u(r) \quad (136\text{a-c})$$

Therefore, the equations of motion reduce to

$$\nabla^4 u - \frac{1}{l^2}\nabla^2 u = -\frac{1}{\eta}\frac{\partial p}{\partial z}, \quad p = p(z) \quad (137\text{a,b})$$

where

$$\nabla^2 u = \frac{1}{r}\frac{\partial}{\partial r}\left(r\frac{\partial u}{\partial r}\right) \quad (138)$$

Since gradient of pressure $\frac{\partial p}{\partial z}$ is constant, the general solution to this equation is given as[8]

$$u = A_0 + A_1 \ln r + \frac{1}{4\mu}\frac{\partial p}{\partial z}r^2 + B_0 I_0\left(\frac{r}{l}\right) + B_1 K_0\left(\frac{r}{l}\right) \quad (139)$$

where $I_0$ and $K_0$ are modified Bessel functions of order zero. In the cylindrical coordinates $(r,\theta,z)$, we have

$$\nabla \times \mathbf{u} = 2\boldsymbol{\omega} = \left(\frac{1}{r}\frac{\partial v_z}{\partial \theta} - \frac{\partial v_\theta}{\partial z}\right)\hat{\mathbf{r}} + \left(\frac{\partial v_r}{\partial z} - \frac{\partial v_z}{\partial r}\right)\hat{\boldsymbol{\theta}} + \frac{1}{r}\left[\frac{\partial}{\partial r}(rv_\theta) - \frac{\partial v_r}{\partial \theta}\right]\hat{\mathbf{z}} \qquad (140)$$

which gives the non-zero component of angular velocity

$$\omega_\theta = -\frac{1}{2}\frac{\partial u}{\partial r} = -\frac{A_1}{2r} - \frac{1}{4\mu}\frac{\partial p}{\partial z}r - \frac{1}{2l}B_0 I_1\!\left(\frac{r}{l}\right) + \frac{1}{2l}B_1 K_1\!\left(\frac{r}{l}\right) \qquad (141)$$

where $I_1$ and $K_1$ are modified Bessel functions of first order. We also have

$$2\mathbf{K} = \nabla \times \boldsymbol{\omega} = \left(\frac{1}{r}\frac{\partial \omega_z}{\partial \theta} - \frac{\partial \omega_\theta}{\partial z}\right)\hat{\mathbf{r}} + \left(\frac{\partial \omega_r}{\partial z} - \frac{\partial \omega_z}{\partial r}\right)\hat{\boldsymbol{\theta}} + \frac{1}{r}\left[\frac{\partial}{\partial r}(r\omega_\theta) - \frac{\partial \omega_r}{\partial \theta}\right]\hat{\mathbf{z}} \qquad (142)$$

which gives the non-zero mean curvature component

$$\mathsf{K}_z = \frac{1}{2r}\frac{\partial (r\omega_\theta)}{\partial r} = -\frac{1}{4\mu}\frac{\partial p}{\partial z} - \frac{1}{4l^2}B_0 I_0\!\left(\frac{r}{l}\right) - \frac{1}{4l^2}B_1 K_0\!\left(\frac{r}{l}\right) \qquad (143)$$

Therefore,

$$M_z = -M_{r\theta} = -8\eta \mathsf{K}_z = -4\eta\left[-\frac{1}{2\mu}\frac{\partial p}{\partial z} - \frac{1}{2l^2}B_0 I_0\!\left(\frac{r}{l}\right) - \frac{1}{2l^2}B_1 K_0\!\left(\frac{r}{l}\right)\right] \qquad (144)$$

The above equations are general for the flow between two concentric pipes. However, for our present single pipe problem, the finiteness of $u$, $\omega_\theta$ and $M_{r\theta}$ at $r = 0$ requires $A_1 = B_1 = 0$. The other boundary conditions are

$$u = 0, \ M_{r\theta} = 0, \text{ at } r = R \qquad (145\text{a,b})$$

Imposing these boundary conditions, we obtain

$$u = -\frac{1}{4\mu}\frac{\partial p}{\partial z}R^2\left[1 - \frac{r^2}{R^2} - 4\frac{l^2}{R^2}\left(1 - \frac{I_0\!\left(\frac{r}{l}\right)}{I_0\!\left(\frac{R}{l}\right)}\right)\right] \qquad (146\text{a})$$

$$\omega_\theta = -\frac{1}{4\mu}\frac{\partial p}{\partial z}R\left[\frac{r}{R} - 2\frac{l}{R}\frac{I_1\!\left(\frac{r}{l}\right)}{I_0\!\left(\frac{R}{l}\right)}\right] \qquad (146\text{b})$$

$$M_{r\theta} = -M_z = -2l^2 \frac{\partial p}{\partial z}\left[1 - \frac{I_0\left(\frac{r}{l}\right)}{I_0\left(\frac{R}{l}\right)}\right] \qquad (146c)$$

Again, the shear stress has two parts, corresponding to the symmetric and skew-symmetric contributions. The symmetric part is

$$T_{(rz)} = \mu \frac{\partial u}{\partial r} = -\frac{\partial p}{\partial z}\left[-\frac{1}{2}r + l\frac{I_1\left(\frac{r}{l}\right)}{I_0\left(\frac{R}{l}\right)}\right] \qquad (147)$$

From the couple-stress, we find

$$\nabla \times \mathbf{M} = \left(\frac{1}{r}\frac{\partial M_z}{\partial \theta} - \frac{\partial M_\theta}{\partial z}\right)\hat{\mathbf{r}} + \left(\frac{\partial M_r}{\partial z} - \frac{\partial M_z}{\partial r}\right)\hat{\boldsymbol{\theta}} + \frac{1}{r}\left[\frac{\partial}{\partial r}(rM_\theta) - \frac{\partial M_r}{\partial \theta}\right]\hat{\mathbf{z}} \qquad (148)$$

which gives

$$\mathbf{S} = \frac{1}{2}\nabla \times \mathbf{M} = -\frac{1}{2}\frac{\partial M_z}{\partial r}\hat{\boldsymbol{\theta}} = l\frac{\partial p}{\partial z}\frac{I_1\left(\frac{r}{l}\right)}{I_0(a)}\hat{\boldsymbol{\theta}} = S_\theta \hat{\boldsymbol{\theta}} \qquad (149)$$

Then, the skew-symmetric part of the shear stress is

$$T_{[rz]} = S_\theta = l\frac{\partial p}{\partial z}\frac{I_1\left(\frac{r}{l}\right)}{I_0\left(\frac{R}{l}\right)} \qquad (150)$$

and for total shear stress $T_{rz}$, we have

$$T_{rz} = \frac{1}{2}\frac{\partial p}{\partial z}r \qquad (151)$$

It is seen that the shear stress in the flow direction $T_{rz}$ is similar to the non-polar case, which gives the wall shear stress as $\frac{1}{2}\frac{\partial p}{\partial z}R$. However, the force-stress tensor is not symmetric in the polar case and we find

$$T_{zr} = \frac{1}{2}\frac{\partial p}{\partial z}r - 2l\frac{\partial p}{\partial z}\frac{I_1\left(\frac{r}{l}\right)}{I_0\left(\frac{R}{l}\right)} \tag{152}$$

The flow rate can be obtained from the relation

$$Q = \int_0^R 2\pi r u \, dr \tag{153}$$

The integration gives

$$\frac{Q}{Q_0} = 1 - 8\frac{l^2}{R^2} + 16\frac{l^4}{R^4}\frac{I_1\left(\frac{R}{l}\right)}{I_0\left(\frac{R}{l}\right)} \tag{154}$$

where

$$Q_0 = -\frac{\pi}{8\mu}\frac{\partial p}{\partial z}R^4 \tag{155}$$

is the classical flow rate, when there are no couple-stresses in the non-polar fluid.

## C. Creeping flow past a sphere

As a third application, we study the effect of couple-stresses on the creeping flow of a polar fluid past a sphere. This problem was first formulated by Stokes[10] within the context of the indeterminate couple-stress theory. Here we follow very closely the presentation given by Stokes[10], but with some deviation due to the current consistent polar fluid formulation.

In creeping flow, the effect of the inertia term, corresponding to acceleration, is neglected. Therefore, the equation of motion for an incompressible polar fluid reduces to

$$\nabla p = \mu \nabla^2 \mathbf{v} - \eta \nabla^2 \nabla^2 \mathbf{v} \tag{156}$$

Here, we use spherical polar coordinates $(r,\theta,\varphi)$, centered on the sphere. We also assume that at large distances from the sphere, the flow is parallel to the $z$-axis corresponding to $\theta = 0$ and that the velocity has the magnitude $U_\infty$. We proceed with a method similar to that used by Stokes[10].

The continuity equation for this axisymmetric flow, where $v_\varphi = 0$, reduces to

$$\frac{1}{r^2}\frac{\partial}{\partial r}(r^2 v_r) + \frac{1}{r\sin\theta}\frac{\partial}{\partial \theta}(v_\theta \sin\theta) = 0 \tag{157}$$

which can be satisfied by the stream function $\psi$, such that

$$v_r = -\frac{1}{r^2 \sin\theta}\frac{\partial \psi}{\partial \theta} \tag{158a}$$

$$v_\theta = \frac{1}{r\sin\theta}\frac{\partial \psi}{\partial r} \tag{158b}$$

In an attempt to satisfy (157), we look for the stream function in the form

$$\psi = f(r)\sin^2\theta \tag{159}$$

where $f$ is a function of $r$, which is yet to be determined. Substituting (159) into (158), the components of the velocity are given by

$$v_r = -\frac{2f(r)}{r^2}\cos\theta \tag{160a}$$

$$v_\theta = \frac{f'(r)}{r}\sin\theta \tag{160b}$$

Following Stokes[10], now we let $\mathbf{w} = \nabla^2 \mathbf{v}$ and $\mathbf{W} = \nabla^2 \mathbf{w} = \nabla^4 \mathbf{v}$. Then, the equation of motion (156) can be written as

$$\frac{1}{\mu}\nabla p = \mathbf{w} - l^2 \mathbf{W} \tag{161}$$

which reduces to the two equations for components

$$\frac{1}{\mu}\frac{\partial p}{\partial r} = w_r - l^2 W_r \tag{162a}$$

$$\frac{1}{\mu}\frac{\partial p}{r\partial \theta} = w_\theta - l^2 W_\theta \tag{162b}$$

By eliminating $p$ between these two equation, we obtain

$$\frac{\partial}{\partial \theta}\left(w_r - l^2 W_r\right) = \frac{\partial}{\partial r}\left(rw_\theta - l^2 rW_\theta\right) \tag{163}$$

Next we consider the components of $\mathbf{w}$. In terms of components of $\mathbf{v}$, the components of $\mathbf{w} = \nabla^2 \mathbf{v}$ are

$$w_r = \nabla^2 v_r - \frac{2}{r^2}\left(\frac{\partial v_\theta}{\partial \theta} + u_r + \cot\theta\, v_\theta\right) = -\frac{2\cos\theta}{r^2}\left(f'' - \frac{2}{r^2}f\right) \tag{164a}$$

$$w_\theta = \nabla^2 v_\theta + \frac{2}{r^2}\left(\frac{\partial v_r}{\partial \theta} - \frac{v_\theta}{2\sin^2\theta}\right) = \frac{\sin\theta}{r}\frac{d}{dr}\left(f'' - \frac{2}{r^2}f\right) \tag{164b}$$

where

$$\nabla^2 \equiv \frac{\partial^2}{\partial r^2} + \frac{2}{r}\frac{\partial}{\partial r} + \frac{1}{r^2}\frac{\partial^2}{\partial \theta^2} + \frac{\cot\theta}{r^2}\frac{\partial}{\partial \theta} \tag{165}$$

One can see that these two equations may be written as[10]

$$w_r = -\frac{2F(r)}{r^2}\cos\theta \tag{166a}$$

$$w_\theta = \frac{F'(r)}{r}\sin\theta \tag{166b}$$

where

$$F(r) = f''(r) - \frac{2}{r^2}f(r) \tag{167}$$

Therefore, for components of $\mathbf{W} = \nabla^2 \mathbf{w}$ in spherical coordinates, we obtain

$$W_r = \nabla^2 w_r - \frac{2}{r^2}\left(\frac{\partial w_\theta}{\partial \theta} + w_r + \cot\theta\, w_\theta\right) = -\frac{2\cos\theta}{r^2}\left(F'' - \frac{2}{r^2}F\right) \tag{168a}$$

$$W_\theta = \nabla^2 u_\theta + \frac{2}{r^2}\left(\frac{\partial u_r}{\partial \theta} - \frac{u_\theta}{2\sin^2\theta}\right) = \frac{\sin\theta}{r}\frac{d}{dr}\left(F'' - \frac{2}{r^2}F\right) \tag{168b}$$

After inserting the expressions (166)-(168) into (163), we obtain the fourth order ordinary differential equation

$$\left(\frac{d^2}{dr^2}-\frac{2}{r^2}\right)^2\left[1-l^2\left(\frac{d^2}{dr^2}-\frac{2}{r^2}\right)\right]f(r)=0 \tag{169}$$

which has the general solution[10]

$$f(r)=A_1 r+A_2 r^2+A_3 r^4+\frac{A_4}{r}+A_5\left(1-\frac{l}{r}\right)e^{\frac{r}{l}}+A_6\left(1+\frac{l}{r}\right)e^{-\frac{r}{l}} \tag{170}$$

The condition that the velocity be finite at infinity requires

$$A_3=A_5=0 \tag{171}$$

and, therefore, we have

$$f(r)=A_1 r+A_2 r^2+\frac{A_4}{r}+A_6\left(1+\frac{l}{r}\right)e^{-\frac{r}{l}} \tag{172a}$$

For the first and second order derivatives of $f$, which are necessary in this analysis, we have

$$f'(r)=A_1+2A_2 r-\frac{A_4}{r^2}-A_6\frac{1}{l}\left(1+\frac{l}{r}+\frac{l^2}{r^2}\right)e^{-\frac{r}{l}} \tag{172b}$$

$$f''(r)=2A_2+\frac{2A_4}{r^3}+A_6\frac{1}{l^2}\left(1+\frac{l}{r}+2\frac{l^2}{r^2}+\frac{2l^3}{r^3}\right)e^{-\frac{r}{l}} \tag{172c}$$

Further, the no slip condition $v_r=v_\theta=0$ on the surface of sphere at $r=R$ implies that $f(R)=f'(R)=0$ and, therefore,

$$A_1 R+A_2 R^2+\frac{A_4}{R}+A_6\left(1+\frac{l}{R}\right)e^{-\frac{R}{l}}=0 \tag{173a}$$

$$A_1+2A_2 R-\frac{A_4}{R^2}-A_6\frac{1}{l}\left(1+\frac{l}{R}+\frac{l^2}{R^2}\right)e^{-\frac{R}{l}}=0 \tag{173b}$$

Also we notice that as $r$ approaches infinity,

$$u_r=U_\infty\cos\theta,\quad u_\theta=-U_\infty\sin\theta \tag{174a,b}$$

Therefore,

$$A_2=-\frac{1}{2}U_\infty \tag{175}$$

We find for vorticity the following relation:

$$\zeta = \nabla \times \mathbf{v} = \frac{1}{r\sin\theta}\left[\frac{\partial}{\partial\theta}(v_\varphi \sin\theta) - \frac{\partial v_\theta}{\partial \varphi}\right]\hat{\mathbf{r}} + \frac{1}{r}\left[\frac{1}{\sin\theta}\frac{\partial v_r}{\partial \varphi} - \frac{\partial}{\partial r}(rv_\varphi)\right]\hat{\boldsymbol{\theta}}$$
$$+ \frac{1}{r}\left[\frac{\partial}{\partial r}(rv_\theta) - \frac{\partial v_r}{\partial \theta}\right]\hat{\boldsymbol{\varphi}} \tag{176}$$

Therefore, with $v_\varphi = 0$, and $\dfrac{\partial}{\partial \varphi} = 0$, the non-zero component of angular velocity is

$$\boldsymbol{\omega} = \omega_\varphi \hat{\boldsymbol{\varphi}} = \frac{1}{2}\left[\frac{1}{r}\frac{\partial}{\partial r}(rv_\theta) - \frac{\partial v_r}{r\partial \theta}\right]\hat{\boldsymbol{\varphi}} \tag{177}$$

where

$$\omega_\varphi = \frac{1}{2}\left[-\frac{2A_1}{r^2} + A_6 \frac{1}{l^3}\left(\frac{l}{r} + \frac{l^2}{r^2}\right)e^{-\frac{r}{l}}\right]\sin\theta \tag{178}$$

We also have

$$2\mathbf{K} = \nabla \times \boldsymbol{\omega} = \frac{1}{r\sin\theta}\left[\frac{\partial}{\partial \theta}(\omega_\varphi \sin\theta) - \frac{\partial \omega_\theta}{\partial \varphi}\right]\hat{\mathbf{r}} + \frac{1}{r}\left[\frac{1}{\sin\theta}\frac{\partial \omega_r}{\partial \varphi} - \frac{\partial}{\partial r}(r\omega_\varphi)\right]\hat{\boldsymbol{\theta}}$$
$$+ \frac{1}{r}\left[\frac{\partial}{\partial r}(r\omega_\theta) - \frac{\partial \omega_r}{\partial \theta}\right]\hat{\boldsymbol{\varphi}} \tag{179}$$

where with $\omega_r = \omega_\theta = 0$, and $\dfrac{\partial}{\partial \varphi} = 0$, this reduces to

$$2\mathbf{K} = \frac{1}{r\sin\theta}\frac{\partial}{\partial \theta}(\omega_\varphi \sin\theta)\hat{\mathbf{r}} - \frac{1}{r}\frac{\partial}{\partial r}(r\omega_\varphi)\hat{\boldsymbol{\theta}} \tag{180}$$

and after substituting, we have

$$2\mathbf{K} = \nabla \times \boldsymbol{\omega} = \left[-\frac{2A_1}{r^3} + A_6 \frac{1}{l^4}\left(\frac{l^2}{r^2} + \frac{l^3}{r^3}\right)e^{-\frac{r}{l}}\right]\cos\theta\,\hat{\mathbf{r}}$$
$$- \left[\frac{A_1}{r^3} - A_6 \frac{1}{2l^4}\left(\frac{l}{r} + \frac{l^2}{r^2} + \frac{l^3}{r^3}\right)e^{-\frac{r}{l}}\right]\sin\theta\,\hat{\boldsymbol{\theta}} \tag{181}$$

Then, for couple-stresses, we have

$$\mathbf{M} = -8\eta\mathbf{K} = -4\eta\nabla \times \boldsymbol{\omega} \tag{182}$$

Therefore, the non-zero components of couple-stress are

$$M_{\varphi\theta} = M_r = -4\eta\left[-\frac{2A_1}{r^3} + A_6\frac{1}{l^4}\left(\frac{l^2}{r^2} + \frac{l^3}{r^3}\right)e^{-\frac{r}{l}}\right]\cos\theta \tag{183a}$$

$$M_{r\varphi} = M_\theta = 4\eta\left[\frac{A_1}{r^3} - A_6\frac{1}{2l^4}\left(\frac{l}{r} + \frac{l^2}{r^2} + \frac{l^3}{r^3}\right)e^{-\frac{r}{l}}\right]\sin\theta \tag{183b}$$

The condition that the tangential couple-stress traction be zero on the surface of the sphere implies that $M_{r\varphi}(R,\theta) = 0$, which requires

$$\frac{A_1}{R^3} - A_6\frac{1}{2l^4}\left(\frac{l}{R} + \frac{l^2}{R^2} + \frac{l^3}{R^3}\right)e^{-\frac{R}{l}} = 0 \tag{184}$$

Therefore, the remainder of the coefficients are found as

$$A_1 = \frac{3}{4}U_\infty R \frac{1 + \frac{l}{R} + \frac{l^2}{R^2}}{1 + \frac{l}{R}} \tag{185a}$$

$$A_4 = -\frac{1}{4}U_\infty R^3 \left[\frac{1 + \frac{l}{R} + 3\frac{l^2}{R^2} + 6\frac{l^3}{R^3} + 6\frac{l^4}{R^4}}{1 + \frac{l}{R}}\right] \tag{185b}$$

$$A_6 = \frac{3}{2}l^2 e^{\frac{R}{l}}U_\infty \frac{1}{1 + \frac{R}{l}} \tag{185c}$$

Therefore, the creeping flow solution can be written

$$v_r = U_\infty\cos\theta - U_\infty\frac{1}{1+\frac{l}{R}}\left\{\begin{array}{l}\frac{3}{2}\left(1+\frac{l}{R}+\frac{l^2}{R^2}\right)\frac{R}{r} \\ -\frac{1}{2}\left(1+\frac{l}{R}+3\frac{l^2}{R^2}+6\frac{l^3}{R^3}+6\frac{l^4}{R^4}\right)\frac{R^3}{r^3} \\ +3\frac{l}{R}e^{\frac{R-l}{l}}\left(\frac{l^2}{r^2}+\frac{l^3}{r^3}\right)\end{array}\right\}\cos\theta \tag{186a}$$

$$v_\theta = -U_\infty \sin\theta + U_\infty \frac{1}{1+\frac{l}{R}} \left\{ \begin{array}{l} \frac{3}{4}\left(1+\frac{l}{R}+\frac{l^2}{R^2}\right)\frac{R}{r} \\ +\frac{1}{4}\left(1+\frac{l}{R}+3\frac{l^2}{R^2}+6\frac{l^3}{R^3}+6\frac{l^4}{R^4}\right)\frac{R^3}{r^3} \\ -\frac{3}{2}\frac{l}{R}e^{\frac{R-r}{l}}\left(\frac{l}{r}+\frac{l^2}{r^2}+\frac{l^3}{r^3}\right) \end{array} \right\} \sin\theta \quad (186b)$$

$$\omega_\varphi = -\frac{3}{4}\frac{U_\infty}{R+l}\left[\left(1+\frac{l}{R}+\frac{l^2}{R^2}\right)\frac{R^2}{r^2} - e^{\frac{R-r}{l}}\left(\frac{l}{r}+\frac{l^2}{r^2}\right)\right]\sin\theta \quad (186c)$$

$$M_r = M_{\phi\theta} = 6\eta U_\infty \frac{1}{R+l}\frac{1}{R}\left[\left(1+\frac{l}{R}+\frac{l^2}{R^2}\right)\frac{R^3}{r^3} - e^{\frac{R-r}{l}}\frac{R}{r}\left(\frac{l}{r}+\frac{l^2}{r^2}\right)\right]\cos\theta \quad (186d)$$

$$M_\theta = M_{r\varphi} = 3\eta U_\infty \frac{1}{R+l}\frac{1}{R}\left[\left(1+\frac{l}{R}+\frac{l^2}{R^2}\right)\frac{R^3}{r^3} - e^{\frac{R-r}{l}}\frac{R}{r}\left(1+\frac{l}{r}+\frac{l^2}{r^2}\right)\right]\sin\theta \quad (186e)$$

With $M_\varphi = 0$ and $\frac{\partial}{\partial\varphi} = 0$, the vector **S** reduces to

$$\mathbf{S} = \frac{1}{2}\nabla\times\mathbf{M} = \frac{1}{2}\frac{1}{r}\left[\frac{\partial}{\partial r}(rM_\theta) - \frac{\partial M_r}{\partial\theta}\right]\hat{\boldsymbol{\varphi}} = S_\varphi\hat{\boldsymbol{\varphi}} \quad (187)$$

Thus,

$$S_\varphi = \frac{3}{2}\mu U_\infty \frac{1}{R+l}e^{\frac{R-r}{l}}\left(\frac{l}{r}+\frac{l^2}{r^2}\right)\sin\theta \quad (188)$$

Therefore, from (188), the skew-symmetric part of $T_{r\theta}$ is

$$T_{[r\theta]} = -S_\varphi = -\frac{3}{2}\mu U_\infty \frac{1}{R+l}e^{\frac{R-r}{l}}\left(\frac{l}{r}+\frac{l^2}{r^2}\right)\sin\theta \quad (189)$$

Interestingly, on the surface of sphere, this becomes

$$T_{[r\theta]}(R,\theta) = -\frac{3}{2}\mu U_\infty \frac{l}{R^2}\sin\theta \quad (190)$$

On the other hand, the symmetric part of $T_{r\theta}$ is given by

$$T_{(r\theta)} = -\frac{3}{2}\mu U_\infty \frac{1}{R+l}\left\{\left(1+\frac{l}{R}+\frac{3l^2}{R^2}+\frac{6l^3}{R^3}+\frac{6l^4}{R^4}\right)\frac{R^4}{r^4} - e^{\frac{R-r}{l}}\left(\frac{l}{r}+\frac{3l^2}{r^2}+\frac{6l^3}{r^3}+\frac{6l^4}{r^4}\right)\right\}\sin\theta \quad (191)$$

which on the surface of sphere becomes

$$T_{(r\theta)}(R,\theta) = -\frac{3}{2}\mu U_\infty \frac{1}{R+l}\sin\theta \quad (192)$$

Therefore, the overall shear stress $T_{r\theta}$ can be written

$$T_{r\theta} = T_{(r\theta)} + T_{[r\theta]} = -\frac{3}{2}\mu U_\infty \frac{1}{R+l}\left\{\left(1+\frac{l}{R}+\frac{3l^2}{R^2}+\frac{6l^3}{R^3}+\frac{6l^4}{R^4}\right)\frac{R^4}{r^4} - e^{\frac{R-r}{l}}\left(\frac{2l^2}{r^2}+\frac{6l^3}{r^3}+\frac{6l^4}{r^4}\right)\right\}\sin\theta \quad (193)$$

which on the surface of the sphere becomes

$$T_{r\theta}(R) = T_{(r\theta)}(R) + T_{[r\theta]}(R) = -\frac{3}{2}\mu U_\infty \frac{1}{R}\frac{1+\frac{l}{R}+\frac{l^2}{R^2}}{1+\frac{l}{R}}\sin\theta \quad (194a)$$

$$T_{\theta r}(R) = T_{(r\theta)}(R) - T_{[r\theta]}(R) = -\frac{3}{2}\mu U_\infty \frac{1}{R}\frac{1-\frac{l}{R}-\frac{l^2}{R^2}}{1+\frac{l}{R}}\sin\theta \quad (194b)$$

Notice that our results here are similar to those given by Stokes[10], if we assume that $\eta' = -\eta$.

Furthermore, the shear drag is given by

$$D_s = \int_0^{2\pi}\int_0^\pi \left[-T_{r\theta}(R)\sin\theta\right]R^2\sin\theta\, d\theta\, d\phi \quad (195)$$

so that

$$D_s = 4\pi\mu R U_\infty \frac{1+\frac{l}{R}+\frac{l^2}{R^2}}{1+\frac{l}{R}} \quad (196)$$

Now we obtain the pressure distribution as follows. First, we note that

$$\frac{1}{\mu}\frac{\partial p}{\partial r} = -\frac{2\cos\theta}{r^2}\left[F(r) - l^2\left(F'' - \frac{2}{r^2}F\right)\right] \quad (197a)$$

$$\frac{1}{\mu}\frac{\partial p}{\partial \theta} = \frac{d}{dr}\left[F - l^2\left(F'' - \frac{2}{r^2}F\right)\right]\sin\theta \tag{197b}$$

Integration of these two equations gives the pressure as

$$p = C - \frac{3}{2}\mu U_\infty R \frac{1 + \frac{l}{R} + \frac{l^2}{R^2}}{1 + \frac{l}{R}} \frac{1}{r^2}\cos\theta \tag{198}$$

where $C$ is a constant. The condition that $p = p_\infty$ at $r = \infty$ implies that $C = p_\infty$. Therefore,

$$p = p_\infty - \frac{3}{2}\mu U_\infty \frac{1}{R+l}\left(1 + \frac{l}{R} + \frac{l^2}{R^2}\right)\frac{R^2}{r^2}\cos\theta \tag{199}$$

It is also seen that the normal radial stress $T_{rr}$ is

$$T_{rr} = -p + 2\mu D_{rr} \tag{200}$$

which gives

$$T_{rr} = -p_\infty + 3\mu U_\infty \frac{1}{R+l}\left\{\begin{array}{l}\frac{3}{2}\left(1 + \frac{l}{R} + \frac{l^2}{R^2}\right)\frac{R^2}{r^2} - \left(1 + \frac{l}{R} + 3\frac{l^2}{R^2} + 6\frac{l^3}{R^3} + 6\frac{l^4}{R^4}\right)\frac{R^4}{r^4} \\ +2e^{\frac{R-r}{l}}\left(\frac{l^2}{r^2} + \frac{3l^3}{r^3} + \frac{3l^4}{r^4}\right)\end{array}\right\}\cos\theta \tag{201}$$

Interestingly, on the surface of the sphere

$$D_{rr}(R,\theta) = 0 \tag{202}$$

which shows

$$T_{rr}(R,\theta) = -p(R) \tag{203}$$

where the pressure on the sphere is

$$p(R,\theta) = p_\infty - \frac{3}{2}\mu U_\infty \frac{1}{R+l}\left(1 + \frac{l}{R} + \frac{l^2}{R^2}\right)\cos\theta \tag{204}$$

Therefore, the pressure drag is given by

$$D_n = \int_0^{2\pi}\int_0^\pi \left[-p(R)\cos\theta\right]R^2\sin\theta\, d\theta\, d\phi \tag{205}$$

which yields

$$D_n = 2\pi\mu R U_\infty \frac{1 + \dfrac{l}{R} + \dfrac{l^2}{R^2}}{1 + \dfrac{l}{R}} \tag{206}$$

Then, from (196) and (206), the total drag $D = D_s + D_n$ is given by

$$D = 6\pi\mu R U_\infty \frac{1 + \dfrac{l}{R} + \dfrac{l^2}{R^2}}{1 + \dfrac{l}{R}} \tag{207}$$

The well-known result for the drag on a sphere for the non-polar case ($l \to 0$) is given by

$$D_0 = 6\pi\mu R U_\infty \tag{208}$$

Therefore, we have

$$\frac{D}{D_0} = \frac{1 + \dfrac{l}{R} + \dfrac{l^2}{R^2}}{1 + \dfrac{l}{R}} \tag{209}$$

All of the fully determinate results presented here for the sphere problem are identical to those of Stokes, if we select $\eta' = -\eta$ in his theory, although this particular case was never considered within the context of his indeterminate theory[10]. For sufficiently small spheres, this example seems to offer a simple experiment for investigating the presence of couple-stresses, a suggestion that was first mentioned by Stokes[10].

## VII. CONCLUDING REMARKS

Mindlin and Tiersten[4] developed the couple stress theory for solid mechanics. Subsequently Stokes[8] extended by analogy the formulation for fluid dynamics. For fluids, this theory is based upon a consistent continuum description of the kinematics with vorticity and curvature rates defined directly from the velocity field, unlike other micropolar theories, which lack such a firm kinematical foundation. However, in these theories, the spherical part of the couple-stress tensor remains indeterminate and body couples appear in the constitutive relations for the force-stress tensor[4]. This, of course, severely hampers the applicability of the theories and brings into doubt the existence of such stress components. In the present paper, we have given careful consideration to the kinematical description, energy relations and boundary conditions to remove all the indeterminacies and inelegant aspects of the formulation. The result is the definition of a fully consistent couple-stress theory for polar continua. It is seen that the couple stress tensor is skew-symmetric and energetically conjugate to the mean curvature rate tensor, which also is skew-symmetrical. Then, in the present work, we have derived the constitutive relations for all of the components of the force stress and couple stress tensors within a linear viscous fluid.

The present continuum theory has the potential to describe matter in micro- and nano-scale and to serve as a bridge to atomistic theories. In his review paper on flows with significant orientational effects, Rae[11] discusses a broad range of applications in which couple-stresses may play an important role. Included are applications in ordered fluids (e.g., liquid crystals), kinetic theory based on intermolecular potentials representing non-central force fields, suspensions with angular momentum exchange, and turbulence. It would be interesting to revisit these topics in light of the present determinate couple-stress theory. The primary impact could be in the field of turbulence research, where a direct connection between Navier-Stokes equations and turbulence phenomena has never been satisfactorily established. The present theory, with its dependence on a natural length scale $l = (\eta/\mu)^{1/2}$ and couple-stresses, which are proportional to the curl of the vorticity, may lead to the development of stronger theoretical linkages.

Beyond this, the present polar theory should be useful for the development of nonlinear fluid mechanics, and also solid mechanics formulations that may govern the behavior of solid continua at the smallest scales.